\documentclass[review]{elsarticle}

\usepackage{lineno,hyperref}
\modulolinenumbers[5]
\usepackage[table,xcdraw]{xcolor}

\journal{Elsevier Internet of Things Journal }

\newcommand{\SF}[1]{\small{\textsf{#1}}}
\begin{document}
%\includepdf[pages={1}]{elsarticle-template.pdf}
\begin{frontmatter}

\title{802.11g Signal Strength Evaluation in an Industrial Environment}%\tnoteref{mytitlenote}}
%\tnotetext[mytitlenote]{Fully documented templates are available in the elsarticle package on \href{http://www.ctan.org/tex-archive/macros/latex/contrib/elsarticle}{CTAN}.}

%% Group authors per affiliation:
\author[mymainaddress,mysecondaryaddress]{Dalton Cézane Gomes Valadares\corref{mycorrespondingauthor}}%\fnref{myfootnote}}
\ead{dalton.valadares@embedded.ufcg.edu.br}
%\fntext[myfootnote]{Since 1880.}

\author[mytertiaryaddress]{Joseana Macêdo Fechine Régis de Araújo}
\ead{joseana@computacao.ufcg.edu.br}

\author[myquaternaryaddress]{Marco Aurélio Spohn}
\ead{marco.spohn@uffs.edu.br}

\author[mysecondaryaddress]{Angelo Perkusich}
\ead{perkusic@dee.ufcg.edu.br}

\author[mysecondaryaddress]{Kyller Costa Gorgônio}
\ead{kyller@computacao.ufcg.edu.br}

\author[mytertiaryaddress]{Elmar Uwe Kurt Melcher}
\ead{elmar@computacao.ufcg.edu.br}

%% or include affiliations in footnotes:

\cortext[mycorrespondingauthor]{Corresponding author}

\address[mymainaddress]{Federal Institute of Pernambuco, Caruaru, PE, Brazil}
\address[mysecondaryaddress]{Embedded Systems and Pervasive Computing Laboratory, Federal University of Campina Grande, Campina Grande, PB, Brazil}
\address[mytertiaryaddress]{Federal University of Campina Grande, Campina Grande, PB, Brazil}
\address[myquaternaryaddress]{Federal University of Fronteira Sul, Chapecó, SC, Brazil}

\begin{abstract}
The advances in wireless network technologies and Industrial Internet of Things (IIoT) devices are easing the establishment of what is called Industry 4.0. For the industrial environments, the wireless networks are more suitable mainly due to their great flexibility, low deployment cost and for being less invasive. Although new wireless protocols are emerging or being updated, changes in existing industries generally can lead to large expenditures. As the well known and accepted IEEE 802.11g standard, mostly used in residential and commercial applications, has a low deployment and maintenance cost, many industries also decide to adopt it. In this scenario, there is a need to evaluate the signal quality to better design the network infrastructure in order to obtain good communication coverage. In this work, we present a practical study about the 802.11g signal strength in a thermoelectric power plant. We collected signal strength values in different points along the engine room and compared our measured values with the estimated ones through the Log-Distance Path Loss model. We concluded that it is possible to use this model in an industrial environment to estimate signal strength with a low error by choosing the right propagation (path loss) exponent.
\end{abstract}


\end{frontmatter}

%\linenumbers

\section{Introduction}
\label{sec:intro}
The technological advances have enabled an increasingly adoption of interconnected devices and applications, in the most diverse areas, such as health care monitoring, vehicle and object tracking, industrial and environmental monitoring. Due to the nature of most of such scenarios, the wireless connectivity is almost a mandatory communication requirement, since it provides a better flexibility and is less invasive than wired technologies considering the devices usually employed (sensors, actuators, embedded systems, etc.).

Many of these applications, besides of the wireless requirement, also require a good level of reliability. For example, Industry 4.0 \cite{Jeschke2017,7883994} and Industrial Internet of Things (IIoT) \cite{BOYES20181}, usually demands some special requirements, because the applications have to monitor machines' operating parameters taking into account environment constraints such as the incidence of magnetic interference. Therefore, it is paramount having a reliable wireless infrastructure in order to allow the responsible sector taking the right decisions in a timely manner.

This kind of application commonly runs on devices with limited processing and storage resources, commonly requiring more powerful devices acting as gateways. This scenario is known as fog computing \cite{valadares18,Maiti19}, when gateways perform some data processing before sending them to a server/cloud, reducing latency and response time of service~\cite{Singh20} since these gateways are closer to devices than cloud servers. To allow a suitable communication network for these devices, it is important to know the behavior of the signal propagation in the specific environment, in order to better plan and deploy the communication infrastructure \cite{chebil13,Pereira18}.

We have conducted a study aimed at investigating the wireless signal propagation in an industrial environment. To this end, we measured the signal strength in 20 points inside the engine room of a Brazilian thermoelectric power plant, located at Northeast region. After gathering the measurements, we employed a path loss model for estimating signal decaying resulting from other sources of interference, path length, the medium, etc. Once we had both real (measured) and estimated (path loss model) signal strength values, we compared them to verify if this model was adequate to predict the signal propagation in an industrial environment.

Although specific protocols, such as Wireless Hart and ZigBee (both based on the IEEE 802.15.4 standard), normally are more suitable for industrial environments, as there was a legacy IEEE 802.11g infrastructure in the thermoelectric power plant and many of the fog devices (\textit{i.e.}, IoT gateways) support communication through this technology, we decided to perform the measurements based on the available  infrastructure. To estimate the propagation loss and investigate its adequacy to industrial environments, we decided to use the Log-Distance Path Loss (LDPL) model, since it is a well accepted model in the literature, and it is also the basis for many other models. The LDPL is applied to indoor and outdoor environments with the presence of obstacles, having a propagation exponent that indicates whether the environment has more or less obstacles, impacting on the computed loss.

Regarding our objective, we have elaborated two  research  questions for guiding this study:
\begin{enumerate}
\item Can the Log-Distance Path Loss model be applied to estimate the signal strength at an industrial environment?
\item What is a good propagation exponent (parameter) to adopt when using the LDPL model at an industrial environment?
\end{enumerate}

This document is an extension of Valadares et al.~\cite{valadares19} %work
%presenting some of the results found in his master thesis~\cite{valadares15}
and is organized as follows: an introduction to path loss models and the model evaluated in this work is briefly given in Section~\ref{sec:background}; in Section~\ref{sec:relworks}, we present some works that also investigate signal propagation; the methodology and experiments are described in Section~\ref{sec:methodology}; the results and a brief discussion are pointed in Section~\ref{sec:results}; lastly, we present the final considerations regarding this presented work and mention some suggestions for future work in Section \ref{sec:conclusion}.
\section{Signal Propagation and Path Loss Models} 
\label{sec:background}
A common effect that occurs when a signal travels through a communication channel is its power level decreasing as the distance increases. Depending on the communication medium as well the paths taken by the signal to reach the destiny, it can also suffer distortion.

A propagation model refers to the way the signal is propagated in the medium, considering effects such as reflection, diffraction, refraction, etc. The signal path loss, or signal power loss, usually occurs with the attenuation of this signal, when there is a reduction in power density. Some of the reasons for this loss / reduction are: reflection, refraction, diffraction, absorption, terrain contours, propagation medium (dry or moist), distance, etc. \cite{ali10, ndzi12}

To quantify the transmitted signal power decreasing along the space propagation, the path loss models are used. To deploy a wireless application, an adequate path loss model is very useful, since it estimates the maximum distance possible to establish successfully communication~\cite{Grant2019PathLM}.

There are several signal propagation/path loss models, some more realistic than others. Some models, such as the Rayleigh fading model, consider the effect of the propagation environment on the signal (when there is no propagation in the line-of-sight, i.e. when there are obstacles); others, such as the Rician fading model, consider that there is a line-of-sight for communication~\cite{ndzi12}.

In visibility conditions, when there is the so-called line-of-sight between the transmitting and receiving antennas, the loss in the link can be considered, simplistically, as corresponding to the loss in free space. The loss in free space is related to the signal energy dispersion along the propagation path and is determined by the Friss Equation~\cite{Shaw2013}, whose power, FSPL (Free Space Path Loss), is calculated from Equation 1.

$$
FSPL(dB) = 20 log(d) + 20 log(f) + 92,44 - Gt – Gr \eqno{(1)}
\label{eq1}
$$

\noindent where: \(d = \) is the distance; \(f = \) is the frequency; \(Gt = \) is the transmitting antenna gain; and \(Gr = \) receiving antenna gain. 

In a scenario with no line-of-sight (NLOS), the path losses between the transmitter and receiver antennas are determined by a more realistic model, which must take into account the most diverse types of obstacles that cause signal attenuation, reflection, refraction, diffraction, etc. Some of these models even consider walls and floors in buildings, either indoors or outdoors.

\subsection{Log-Distance Path Loss Model}
In this section, we present a brief description about the Log-Distance Path Loss model, which is the chosen propagation model to be evaluated in our industrial environment.

A simple and well accepted propagation model, which takes into account the existence of some obstacles, in open and closed environments, is the Log Distance Path Loss \cite{ali10,FariaModelingSA}, whose path loss is calculated with Equation 2: 
$$
L(d) = L_0 + 10 n log(d) \eqno{(2)}
\label{eq2}
$$
\noindent where: \(d = \) is the distance; \(L_0 = \) is the signal strength from 1m of the transmitter (antenna); and \(n = \) propagation exponent (depends of the obstacles in the environment;

Many models derive from LDPL, with adjustments in the propagation exponent and addition of parameters related to the number of walls, floors, etc. In general terms, the propagation exponent varies according the environment, as described below \cite{srinivasa09}:
\begin{itemize}
\item Free space - 2;
\item Cellular radio in urban area - 2.7 to 3.5;
\item Cellular radio in urban area with fading - 3 to 5;
\item Closed environment with line of sight - 1.6 to 1.8;
\item Building with obstacles - 4 to 6;
\item Factory with obstacles - 2 to 3.
\end{itemize}

\section{Related Works}
\label{sec:relworks}
%In this section we present some works related to signal strength estimation, mainly with the use of LDPL model.

Faria \cite{FariaModelingSA} carried out a study on the modeling of signal attenuation in 802.11b networks. This study was performed considering internal and external communications, in a building of the Stanford University. The experiment considered 41 measurement points, varying the distances between 1 and 50 meters. The estimated values were calculated with the LDPL model considering a variation in the propagation exponent. The results validated the use of this model, after comparison with measured values, with 4.02 and 3.32 as propagation exponents for internal and external communications, respectively. A similar study was accomplished at Kuala Lumpur University, by Ali \textit{et al}.
\cite{ali10}. They investigated the 802.11g signal propagation in a closed environment and also used the LDPL model to estimate values. The measured values were compared with the estimated ones and the determined propagation exponents, according to floors of the building, were 2.68, 3.22 and 4.

Lkhagvatseren and Hruska \cite{lkhagvatseren11} compared the various path loss models applied to indoor environments, varying different power levels and frequencies, and observing the impact of environmental factors. The results showed that the LDPL model presents good approximation when compared to real values measured. Cheffena and Mohamed \cite{cheffena17} investigated the path loss effects on a wireless sensor network (WSN) in a snowy environment. They measured the path loss at different heights from the ground and compared the results with the values estimated by the "Two-Ray" and "Ray tracing" models. Since the comparison presented significant difference, they proposed new empirical models based on LDPL, which demonstrated good accuracy to apply to WSN deployments in snowy environments.

Japertas \textit{et al.} \cite{japertas12} verified the 802.11g/n signal propagation considering scenarios non-line-of-sight (NLOS) and with line of sight (LOS) inside a building with multiple divisions (obstacles). The measured values were compared to estimated values with LDPL and Free Space Path Loss models. As a result, a new model was proposed, considering NLOS environments with signal transmission along homogeneous walls. Fernández \textit{et al.} \cite{fernandez} proposed some adjustments to LDPL model when applied to digital TV signal transmission. Three signal strength measurement sets, obtained in Lima (Peru), were compared to values estimated with LDPL, Okumura-Hata and ITU-R models. The two adjusted values for propagation exponent were 4.5 and 4.7, providing to LDPL model good approximation to the real values.

Rath \textit{et al.}\cite{rath17} observed that the traditional indoor path loss models are not suitable to Indian scenarios due to some buildings characteristics such as used materials, floor plans, etc. They proposed a non-deterministic statistical path loss model, which was compared with LDPL and another model, becoming a good possibility to be used in India. In an urban environment, in Valencia (Spain), a path loss characterization of the vehicular-to-infrastructure (V2I) channel was carried out, by Rubio \textit{et al.} \cite{rubio15}, also based on LDPL model. The authors investigated the correlation between the height of the antenna used and the propagation (path loss) exponent, varying both values, and concluded that there is no significant correlation degree.

Damsaz~\cite{damsaz17} \textit{et al.} analyzed some wireless propagation characteristics in industrial environments to propose path loss models, focusing on ZigBee technology. The measurements were collected at various factory of factory-like environments, such as a machine shop or an automotive assembly plant. They determined the propagation loss based on the distance, the shadowing level and the RMS delay
spread of the channel. A performance of ZigBee radio using the channels considered was presented. Karaagac \textit{et al.} \cite{karaagac17} also used the LDPL model as a basis to estimate the path loss in an industrial environment. They considered the 802.15.4e technology, proposing three reliable and flexible architectures with 2.4 GHz and 868 MHz.

Ulusar et al.~\cite{Ulusar2019} performed practical tests to assess the performability of a ZigBee network, aiming at an accurate positioning of the sensor nodes. The authors measured the ToF (Time of Flight), which specifies the spent time to propagate the signal between the transmitter and receiver nodes, as well as the RSSI (Received Signal Strength Indicator) and its quality. The results indicate that the positioning based on the ToF measurements is more accurate than based on the received signal strength. Miao et al.~\cite{miao18} conducted experiments in a wheat field performing range measurements in a 2.4 GHz ZigBee link to estimate better accurate locations. Through experimentation in the wheat field and simulation in Matlab, the authors measured the RSSI in difference distances and compared these values with the estimated with three path loss models. The proposed Optimal Fitting Parametric Exponential Decay Model (OFPEDM) presented a better estimation accuracy than the two other path loss models (Free Space Path Loss and Log-Normal Shadowing Model), presenting acceptable errors (0.0004 m to 5.1739 m) when considering wide areas.

Phunthawornwong et al.~\cite{Phunthawornwong18} proposed a location estimated method, using the LDPL model and RSSI measurement data, for indoor environment and location tracking of little robots used in industry and medical applications. The authors adjusted the LDPL model varying the propagation exponent ($n$) and comparing the estimated values with the measured RSSI values obtained through experimentation considering a 2.45 Ghz operation frequency. The results conclude that the LDPL model presents an acceptable average error of 7.42\% by setting the propagation exponent to 3.2. Dharmadhikari et al.~\cite{Dharmadhikari18} also presented a study on the wireless sensor nodes positioning, in a 2.4 GHz ZigBee indoor link, using the LDPL model and RSSI measurements data. The results conclude that the propagation exponent presenting the lower errors is $n = 4.2$. Kun et al.~\cite{Kun19}

A similar analysis, for an outdoor environment with high average temperature (24-50 ºC) and humidity (75-90\%), was carried out by Alhammadi et al.~\cite{Alhammadi19}, considering a 2.4 GHz 802.11 network. The authors also measured the signal strength alogn the distance and compared these values with the estimated with the LDPL model. The results demonstrate that the better estimation was obtained with the propagation exponent $n=2.264$. Pereira et al.~\cite{Pereira18} investigated the signal propagation of a 2.4 GHz ZigBee network in two different environments: an office and a hydroelectric power plant. The RSSI was also measured for different points in both environments and these values were compared with the LDPL model estimated ones. Good results were obtained using the propagation exponent $n$ between 4.5 and 4.6, for the office, and between 5.2 and 6.5 for the power plant. Kun et al.~\cite{Kun19} also performed a signal strength analysis in an industrial factory, for Industrial IoT applications deployment, investigating the signal propagation in two scenarios (LOS and NLOS) regarding the following frequencies: 1.1 GHz, 1.6 GHz, 2.55 GHz and 3.5 GHz. The authors estimated the values with a dual-slope path loss model and compared these values with the obtained measurements. They conclude that the path loss exponents with better estimations, considering both tested scenarios, are 2.7 and 3.3.

As could be seen, many works have concerned with wireless signal propagation, always aiming at a better network infrastructure planning and deployment. The Internet of Things, the Industry 4.0 and the Wireless Sensor Networks increase the importance of this kind of study applied to industrial environments, what reinforces the motivation and justification of our work.

\section{Methodology and Experiment}
\label{sec:methodology}
In this section, we present the experimental design and the methodology employed to carry out this work, describing the environment, the tools, and the performed procedures. As mentioned before, the study considers an IEEE 802.11g infrastructure, since it was already deployed at the thermoelectric power plant.

To help and guide the achievement of a satisfactory answer for our research question, we have defined the business and technical problems, as follow:
\begin{itemize}
\item \textbf{Business problem}: investigate if the LDPL model can be applied reasonably to estimate the signal strength in an industrial environment.
  \item \textbf{Technical problems}: (i) measure the signal strength throughout different points in the engine room; and, compute the signal strength estimated values with the LDPL model and compare them with the measured ones.
\end{itemize}

\subsection{Environment}
\begin{figure}[htpb]
\centering
\includegraphics[width=1\textwidth]{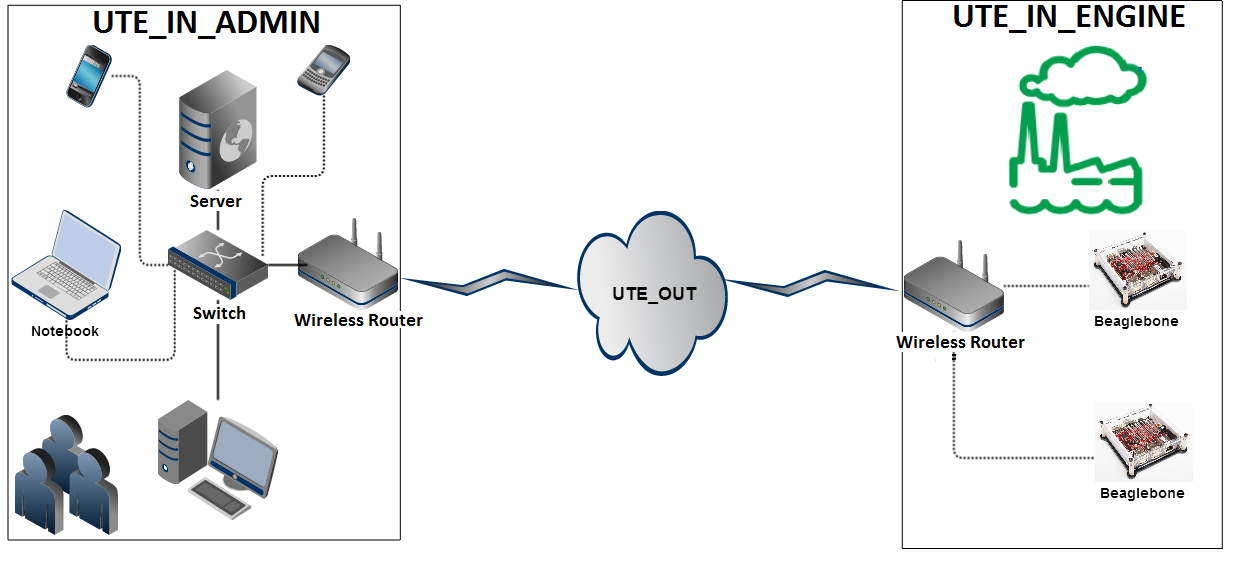}
\caption{Thermoelectric power plant network architecture}
\label{fig:arquitetura}
\end{figure}

The network architecture deployed in the thermoelectric power plant is composed of four wireless communication links, located between the administrative room and the engine room, as shown in Fig.~\ref{fig:arquitetura}. 

The \SF{UTE\_IN\_ADMIN} link is located inside the administrative room, whereas the \SF{UTE\_IN\_ENGINE} is located inside the engine room. The \SF{UTE\_OUT} link connects both \SF{UTE\_IN\_ADMIN} and \SF{UTE\_IN\_ENGINE} links, and it is  equipped with two INTELBRAS WOG 212 antennas, which operate in a frequency of 2.4 GHz, with throughput up to 150 Mbps, compatible with the IEEE 802.11b/g standards, and an integrated antenna with a gain of 12 dBi and a nominal power of 27dBm \cite{intelbras}. Both \SF{UTE\_IN\_ADMIN} and \SF{UTE\_IN\_ENGINE} are characterized by the existence of a TP Link access point, with effective radiated power of 20dB. For our study, we just considered the TP Link access point inside the engine room, which is responsible for the \SF{UTE\_IN\_ENGINE} network.

\subsection{Scenario}

\begin{figure}[ht]
\centering
\includegraphics[width=1\textwidth]{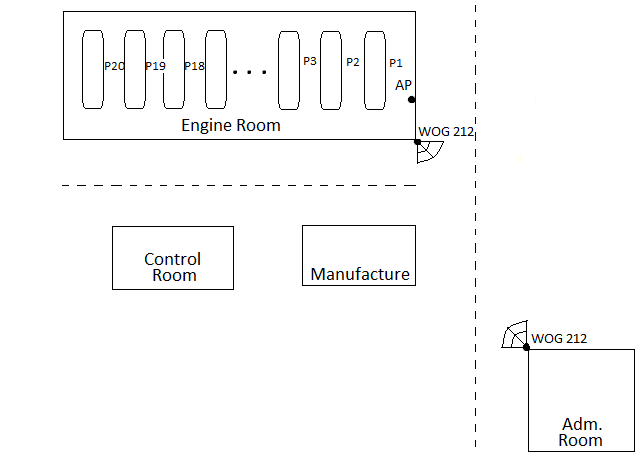}
\caption{Buildings in the power plant and points inside the engine room.}
\label{fig:scenario}
\end{figure}

The physical scenario where the tests were performed is shown in Fig. \ref{fig:scenario}. There are two WOG 212, responsible for the \SF{UTE\_OUT} link, and the AP point is representing the approximate place where TP Link access point is deployed inside the engine room. The distance between two WOG 212 is about 150 m. The engine room is approximately 150 m long, and has 20 engines. 

 The signal strength measurements were performed in distinct points inside the engine room, represented by \SF{P1,$\cdots$, P20} in Fig. \ref{fig:scenario}. These points were chosen to enable the observation of the possible impact of the electromagnetic interference  in the data transmission. They were positioned at the center of the engines, with 6m between consecutive points, except between engines \SF{P10} and \SF{P11}, which were about 12 m from each other.

\subsection{Instrumentation and Tests}
To carry out the tests (i.e., obtaining the signal strength in each point), a notebook running Linux Ubuntu 14 was used. The operating system had to run from pen drive due to electromagnetic interference in the engine room causing failures/crashes to hard drives (HDs), as reported by some engineers of the power plant.

As the main tools, we used \textbf{iwconfig} and \textbf{linssid}. The \textbf{iwconfig} is a Linux tool for the configuration of wireless network interfaces, enabling the setting up and verification of parameters such as channel frequency and signal strength \cite{iwconfig}. The \textbf{linssid} is a wireless network tracker for Linux, by graphically representing sensed networks with the corresponding channels in the frequency domain,  including also the following information: signal quality, signal strength, network ID, channel, and the noise level \cite{linssid}. 

To gather the signal strength and quality, the notebook was positioned on the ground in each defined point, while both tools were employed. The tools collected the data during 40 min for each defined position. To validate our experiment, this process was repeated five times, in different days and shifts. The data obtained through this procedure were eventually validated given that they presented a high degree of similarity.

\section{Results and Discussion}
\label{sec:results}
In this section we present the results and some discussion on that. With the collected data, it was possible to compute the mean value for each point, allowing the comparison of such representative unit with the signal strength estimated with the LDPL model, while varying the propagation exponent ($n$) from 3 to 6. The estimation was calculated by subtracting the value of the path loss computation at each point (obtained with the LDPL model) from the effective power radiated by the access point inside the engine room (20dB). To calculate the loss at each point, a reference loss of -20dB was considered since 1m from the access point, based on the work by Faria~\cite{FariaModelingSA}, which presents an access point with the same characteristics of ours and the same reference distance. 

As explained in the end of the previous Section (\ref{sec:methodology}), we performed the experiment 5 times, in different days and shifts, collecting signal strength data for each determined point. The means of the collected signal strength values for each point and for each test execution are showed in the Table \ref{tab:signalstrength}. Besides, the Table \ref{tab:signalstrength} also contains the estimated signal strength values calculated when considering the path loss results from the LDPL model, as well as the mean signal strength for each point calculated with the mean values of the 5 tests. LDPL\(n\) represents the signal strength values obtained by using the LDPL model with \(n\) as the propagation exponent (3, 4, 5 and 6). Below each point, in the first column, there is the distance, in meters (m), from the point to the access point.

% Please add the following required packages to your document preamble:
% \usepackage{graphicx}
% \usepackage[table,xcdraw]{xcolor}
% If you use beamer only pass "xcolor=table" option, i.e. \documentclass[xcolor=table]{beamer}
\begin{table}[]
\caption{Real and estimated signal strength along the engine room.}
\label{tab:signalstrength}
\centering
\resizebox{\textwidth}{!}{%
\begin{tabular}{ccccccccccc}
\hline
\rowcolor[HTML]{EFEFEF} 
\textbf{\begin{tabular}[c]{@{}c@{}}Point\\ (m)\end{tabular}} & \textbf{\begin{tabular}[c]{@{}c@{}}1º test\\ (dB)\end{tabular}} & \textbf{\begin{tabular}[c]{@{}c@{}}2º test\\ (dB)\end{tabular}} & \textbf{\begin{tabular}[c]{@{}c@{}}3º test\\ (dB)\end{tabular}} & \textbf{\begin{tabular}[c]{@{}c@{}}4º test\\ (dB)\end{tabular}} & \textbf{\begin{tabular}[c]{@{}c@{}}5º test\\ (dB)\end{tabular}} & \textbf{\begin{tabular}[c]{@{}c@{}}Mean\\ (dB)\end{tabular}} & \textbf{\begin{tabular}[c]{@{}c@{}}LDPL3\\ (dB)\end{tabular}} & \textbf{\begin{tabular}[c]{@{}c@{}}LDPL4\\ (dB)\end{tabular}} & \textbf{\begin{tabular}[c]{@{}c@{}}LDPL5\\ (dB)\end{tabular}} & \textbf{\begin{tabular}[c]{@{}c@{}}LDPL6\\ (dB)\end{tabular}} \\ \hline
\begin{tabular}[c]{@{}c@{}}P1\\ (7)\end{tabular} & -45 & -43 & -46 & -45 & -47 & -45,2 & -25,35 & -33,80 & -42,26 & -50,71 \\ \hline
\begin{tabular}[c]{@{}c@{}}P2\\ (13)\end{tabular} & -47 & -53 & -54 & -54 & -59 & -53,4 & -33,42 & -44,56 & -55,70 & -66,84 \\ \hline
\begin{tabular}[c]{@{}c@{}}P3\\ (19)\end{tabular} & -56 & -55 & -55 & -57 & -57 & -56 & -38,36 & -51,15 & -63,94 & -76,73 \\ \hline
\begin{tabular}[c]{@{}c@{}}P4\\ (25)\end{tabular} & -61 & -58 & -55 & -57 & -62 & -58,6 & -41,94 & -55,92 & -69,90 & -83,88 \\ \hline
\begin{tabular}[c]{@{}c@{}}P5\\ (31)\end{tabular} & -63 & -60 & -58 & -61 & -61 & -60,6 & -44,74 & -59,66 & -74,57 & -89,48 \\ \hline
\begin{tabular}[c]{@{}c@{}}P6 \\ (42)\end{tabular} & -66 & -57 & -56 & -57 & -56 & -58,4 & -48,70 & -64,93 & -81,16 & -97,40 \\ \hline
\begin{tabular}[c]{@{}c@{}}P7\\ (48)\end{tabular} & -61 & -58 & -64 & -64 & -65 & -62,4 & -50,44 & -67,25 & -84,06 & -100,87 \\ \hline
\begin{tabular}[c]{@{}c@{}}P8\\ (54)\end{tabular} & -66 & -64 & -65 & -66 & -66 & -65,4 & -51,97 & -69,30 & -86,62 & -103,94 \\ \hline
\begin{tabular}[c]{@{}c@{}}P9\\ (60)\end{tabular} & -68 & -63 & -64 & -68 & -68 & -66,2 & -53,35 & -71,13 & -88,91 & -106,69 \\ \hline
\begin{tabular}[c]{@{}c@{}}P10\\ (66)\end{tabular} & -73 & -69 & -67 & -68 & -68 & -69 & -54,59 & -72,78 & -90,98 & -109,17 \\ \hline
\begin{tabular}[c]{@{}c@{}}P11\\ (78)\end{tabular} & -73 & -62 & -63 & -70 & -69 & -67,4 & -56,76 & -75,68 & -94,61 & -113,53 \\ \hline
\begin{tabular}[c]{@{}c@{}}P12\\ (84)\end{tabular} & -76 & -67 & -67 & -75 & -75 & -72 & -57,73 & -76,97 & -96,21 & -115,46 \\ \hline
\begin{tabular}[c]{@{}c@{}}P13\\ (90)\end{tabular} & -83 & -76 & -74 & -77 & -74 & -76,8 & -58,63 & -78,17 & -97,71 & -117,26 \\ \hline
\begin{tabular}[c]{@{}c@{}}P14\\ (96)\end{tabular} & -82 & -76 & -76 & -78 & -77 & -77,8 & -59,47 & -79,29 & -99,11 & -118,94 \\ \hline
\begin{tabular}[c]{@{}c@{}}P15\\ (102)\end{tabular} & -83 & -78 & -77 & -79 & -78 & -79 & -60,26 & -80,34 & -100,43 & -120,52 \\ \hline
\begin{tabular}[c]{@{}c@{}}P16\\ (113)\end{tabular} & -79 & -79 & -78 & -74 & -72 & -76,4 & -61,59 & -82,12 & -102,65 & -123,19 \\ \hline
\begin{tabular}[c]{@{}c@{}}P17\\ (119)\end{tabular} & -88 & -77 & -73 & -84 & -80 & -80,4 & -62,27 & -83,02 & -103,78 & -124,53 \\ \hline
\begin{tabular}[c]{@{}c@{}}P18\\ (125)\end{tabular} & -86 & -84 & -80 & -84 & -81 & -83 & -62,91 & -83,88 & -104,85 & -125,82 \\ \hline
\begin{tabular}[c]{@{}c@{}}P19\\ (131)\end{tabular} & -89 & -88 & -82 & -84 & -83 & -85,2 & -63,52 & -84,69 & -105,86 & -127,04 \\ \hline
\begin{tabular}[c]{@{}c@{}}P20\\ (137)\end{tabular} & -87 & -84 & -85 & -82 & -83 & -84,2 & -64,10 & -85,47 & -106,84 & -128,20 \\ \hline
\end{tabular}%
}
\end{table}

\begin{figure}[ht]
\centering
\includegraphics[width=\textwidth]{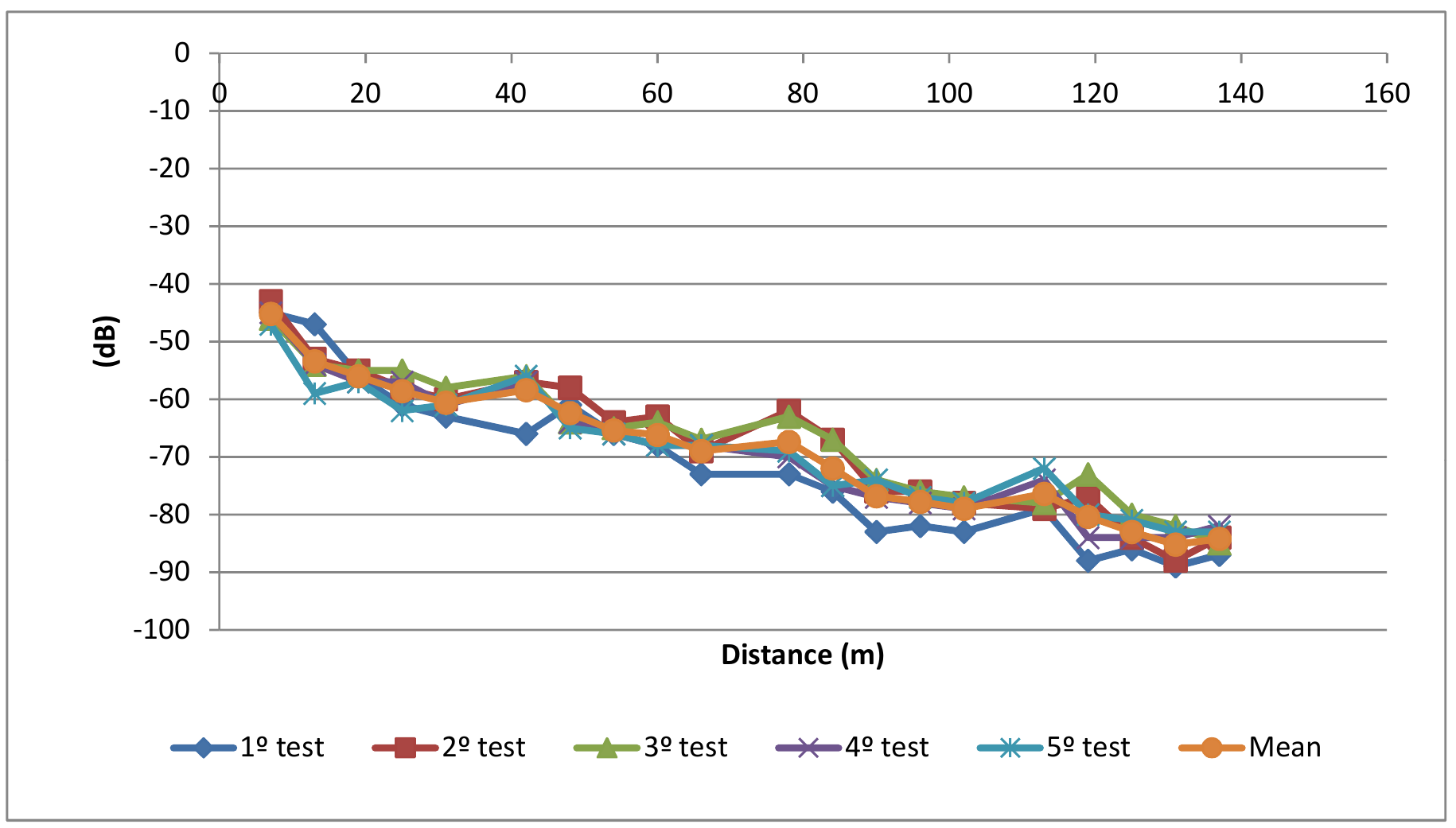}
\vspace{-3em}
\caption{Signal strength for each test and the mean}
\label{fig:sigstren5tests}
\end{figure}

\begin{figure}[!ht]
\centering
\includegraphics[width=\textwidth]{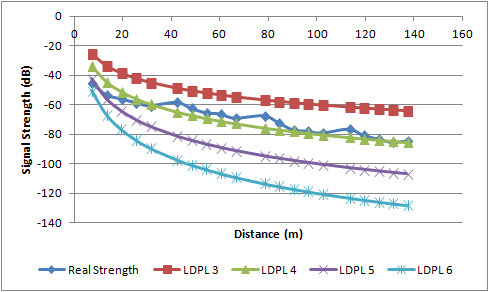}
\vspace{-3em}
\caption{Real and estimated signal strength along the engine room.}
\label{fig:realcurve}
\end{figure}

The signal strength means for each point in each test, as well as the mean value considering all the 5 means, are presented in Figure \ref{fig:sigstren5tests}. As can be seen, both tests collected similar signal strengths for each point. In Figure \ref{fig:realcurve}, we can see the real signal strength (mean of the 5 measurements) along with the estimated ones (considering path loss results with the LDPL model for each propagation exponent value: 3, 4, 5, and 6).

\begin{figure}[!ht]
\centering
\includegraphics[width=\textwidth]{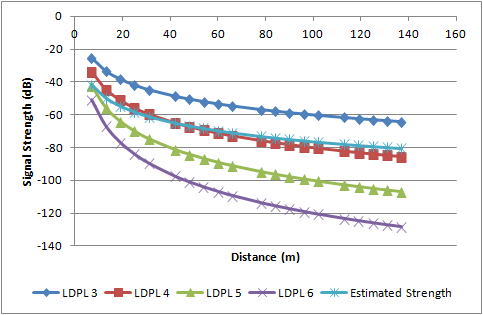} \\
\includegraphics[width=\textwidth]{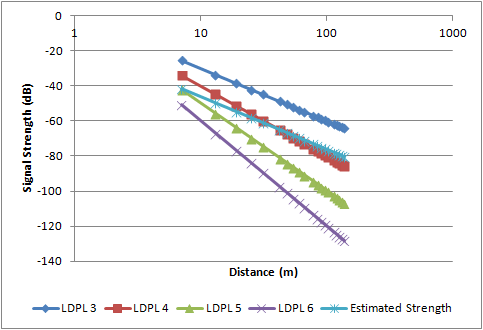}
\caption{Logarithmic regression and LDPL estimated signal strength along the engine room.}
\label{fig:reglogcurve}
\end{figure}

To facilitate the visualization and interpretation of the graphs, a non-linear logarithmic regression was performed, obtaining a coefficient of determination 0.91 (\(R^2 = 0.91\)). In Figure \ref{fig:reglogcurve}, we show the graphs comparing the logarithmic regression with the estimated values in natural (above) and logarithmic scale (below).

As can be seen in Figures \ref{fig:realcurve} and \ref{fig:reglogcurve}, the estimated values closest to the real ones are those obtained with the propagation exponent 4 (i.e., LDPL 4). We calculated the relative error between the real values and the estimated values with LDPL 4, as well as between the logarithmic regression values and the estimated ones with LDPL 4. The relative errors calculated for each point can be seen in the Table \ref{fig:errors}.

% Please add the following required packages to your document preamble:
% \usepackage{graphicx}
% \usepackage[table,xcdraw]{xcolor}
% If you use beamer only pass "xcolor=table" option, i.e. \documentclass[xcolor=table]{beamer}
\begin{table}[htb]
\caption{Relative errors between measured, estimated (LDPL 4) and regression values.}
\label{fig:errors}
\centering
\begin{tabular}{
>{\columncolor[HTML]{EFEFEF}}c c
>{\columncolor[HTML]{EFEFEF}}c c|
>{\columncolor[HTML]{EFEFEF}}c c
>{\columncolor[HTML]{EFEFEF}}c c}
\hline
\multicolumn{4}{c|}{\cellcolor[HTML]{EFEFEF}\textbf{Real x LDPL 4}} & \multicolumn{4}{c}{\cellcolor[HTML]{EFEFEF}\textbf{Regression x LDPL 4}} \\ \hline
\textbf{P1} & 25\% & \textbf{P11} & 12\% & \textbf{P1} & 19\% & \textbf{P11} & 3\% \\ \hline
\textbf{P2} & 17\% & \textbf{P12} & 7\% & \textbf{P2} & 11\% & \textbf{P12} & 4\% \\ \hline
\textbf{P3} & 9\% & \textbf{P13} & 2\% & \textbf{P3} & 7\% & \textbf{P13} & 4\% \\ \hline
\textbf{P4} & 5\% & \textbf{P14} & 2\% & \textbf{P4} & 4\% & \textbf{P14} & 5\% \\ \hline
\textbf{P5} & 2\% & \textbf{P15} & 2\% & \textbf{P5} & 2\% & \textbf{P15} & 5\% \\ \hline
\textbf{P6} & 11\% & \textbf{P16} & 7\% & \textbf{P6} & 0\% & \textbf{P16} & 5\% \\ \hline
\textbf{P7} & 8\% & \textbf{P17} & 3\% & \textbf{P7} & 1\% & \textbf{P17} & 6\% \\ \hline
\textbf{P8} & 6\% & \textbf{P18} & 1\% & \textbf{P8} & 1\% & \textbf{P18} & 6\% \\ \hline
\textbf{P9} & 7\% & \textbf{P19} & 1\% & \textbf{P9} & 2\% & \textbf{P19} & 6\% \\ \hline
\textbf{P10} & 5\% & \textbf{P20} & 2\% & \textbf{P10} & 3\% & \textbf{P20} & 6\% \\ \hline
\end{tabular}
\end{table}

With the exception of some discrepancies for a few points (\(P1\) and \(P2\)), relative errors are all below 15\%, and they are even lower when considering the logarithmic regression. We noticed that the engine room walls worked as a kind of shielding, helping to keep the signal quality, and attenuating the electromagnetic interferences effect on the signal strength. A possible explanation for the discrepancies is that, during the data collection, workers were accessing the engine room for performing their regular activities, and as the doors were opened the shielding effect appeared to be somehow broken, by allowing some signal leakage while impacting the overall signal quality. This makes the results more realistic, since the data collection was carried out in real working shifts, and being the least possible invasive.

As the employed model was not designed to take into account electromagnetic interferences, we are inducted to conclude that the 802.11g signal quality, while considering the power plant environment (characteristics and "shielding"), is not much affected by the interferences from the motors, since the estimated power computed with  propagation exponent 4 (which is used to estimate the path loss in a regular environment with the presence of several obstacles) showed a good approximation with the measured power. In other words, it seems that the existing shielding in the engine room counterbalances the negative effect of electromagnetic interference, characterizing the network behavior as that of a common environment, in which the signal propagation is, many times, more reduced by distance and obstacles.

Thus, the main contributions of this work are: 
\begin{itemize}
\item A practical evaluation in a real industrial environment (i.e., thermoelectric power plant), while similar studies/works usually simulate the local;
\item The determination of the propagation exponent to be considered for 802.11g signal strength estimation in industrial environments (achieving an error bellow 15\% for most of the spots in our evaluation).
\end{itemize}

%%-13,03*LN(B2)-16,369
\section{Conclusion}
\label{sec:conclusion}
%The growth of IoT applications adoption in industrial environments, also known as Industrial IoT applications, awakes some care related to wireless communications infrastructure, as seen earlier in this document. In this sense, it is important to make a good planning, in order to establish the right places to locate the access points so that they can provide the best coverage.

In this paper we presented an evaluation of the 802.11g signal strength in a real industrial environment: a thermoelectric power plant. After an established experimental design, we have collected signal strength measurements along the engine room to compare with the calculated ones by the LDPL model. The performed evaluation, resulted from the calculated values and comparisons, demonstrates that the LDPL model can be used to estimate the 802.11g signal strength in an industrial environment with an error lower than 15\% with respect to most of the points considered. This is possible by setting the propagation exponent \(n\) to 4.

In the context of the experiments and the analysis of the results we concluded that the LDPL model also can be used in industries, assisting in the establishment of a good network infrastructure inside engine rooms with similar characteristics as that we investigated, in a thermal power plant. 

As future work, we consider the application of machine learning techniques to determine a model that better fits to the real path loss in each point, resulting also in a better adjusting to the measured values of signal strength, obtained in the engine room. Furthermore, this experiment can also be replicated to observe the behavior of signal strength from another wireless technologies, such as IEEE 802.11n, IEEE 802.11af and IEEE 802.11ah. Yet, together with the applied path loss model, we suggest the new analyses can be improved by the addition of some heat map tool, identifying points with better coverage and points that are more affected by electromagnetic interference.

\section*{Acknowledgements}
The authors thank Borborema Energ\'etica S.A. and Maracana\'u Geradora de Energia S.A., sponsors of the ANEEL GASIS R\&D project, in which this research was carried out, as well as CNPq, for having financed some months of the master's research of the main author.

\bibliography{elsarticle-template}

\end{document}